# Size-selected polyynes synthesised by submerged arc discharge in water

Sonia Peggiani[1], Andrea Senis[1], Anna Facibeni[1], Alberto Milani[1], Patrick Serafini[1], Gianmarco Cerrato[1], Andrea Lucotti[2], Matteo Tommasini[2], Daniele Fazzi[3], Chiara Castiglioni[2], Valeria Russo[1], Andrea Li Bassi[1] and Carlo S. Casari[1, *]

[1] Micro- and Nanostructured Materials Laboratory, Department of Energy, Politecnico di Milano, via Ponzio 34/3, 20133 Milano, Italy

[2] Department of Chemistry, Materials, and Chemical Engineering "Giulio Natta", Politecnico di Milano, Piazza Leonardo da Vinci 32, 20133 Milano, Italy

[3] Institut für Physikalische Chemie, Department Chemie, Universität zu Köln, Luxemburger Str. 116, D-50939 Köln, Germany

**Abstract**

Polyynes, linear sp-carbon chains of finite length, can be synthesised by submerged arc discharge in liquid, which so far, has been mainly exploited in organic solvents.

In this work, we investigated the technique in water as a cheap and non-toxic solvent to produce polyynes. After optimisation of the process parameters, hydrogen-terminated polyynes ($C_nH_2$: n=6-16) were identified by high-performance liquid chromatography with the support of theoretical calculations. Size-selected polyynes were separately analysed by surface-enhanced Raman spectroscopy allowing to assign the bands of mixed polyynes solution to specific wire lengths. Stabilisation strategies for hydrogen-capped polyynes were also studied, obtaining promising results.

1. **Introduction**

In the last 30 years carbon-based nanomaterials have stimulated an increasing interest in the scientific community [1]. Besides diamond and graphite, characterised by $sp^3$ and $sp^2$ hybridised carbon atoms respectively, carbyne is the one-dimensional infinite wire based on sp-hybridised carbon showing intriguing properties as predicted by theoretical calculations (e.g., high thermal conduction, high electron mobility)[2-8]. The two isomeric configurations of carbyne are cumulene (i.e., sequence of double bonds with metallic behaviour) and polyynes (i.e., sequence of alternated single and triple bonds with semiconducting behaviour).

*Corresponding author. Email: carlo.casari@polimi.it (Carlo Casari)
Phone number: +390223996331

Carbyne-like nanostructures with finite size and specific termination (i.e. end group) are carbon atomic wires (CAWs) appealing for their structure-dependent opto-electronic properties, such as length dependent band gap and tuneable insulator/semiconductor/metallic behaviour. The exploitation of these carbon nanomaterials in technological applications requires the control of their synthesis process and stability [9]. In fact, stability is mainly limited by the occurrence of cross-linking reactions usually leading to reorganisation from sp to $sp^2$ amorphous carbon [10].

Many different techniques can be employed to synthesise sp-carbon systems, from chemical methods, based on polymerisation reactions [11, 12] or on rational synthesis [13, 14], to physical methods, based on quenching of carbon vapour/plasma in gases or liquids[15-19]. The exploitation of a liquid medium to quench a carbon plasma has been proven to be a viable way for the formation of hydrogen-capped polyynes, as in the case of pulsed laser ablation of graphite or fullerene particles suspended in organic solvents [18, 20]. F. Cataldo explored the use of an arc discharge between graphite electrodes submerged in a solvent to produce polyynes [12, 21]. Recently, Wu and co-workers exploited arc discharge between copper electrodes in alkanes to investigate the solvent role in polyyne formation in the absence of carbon species coming from the electrodes [22]. Such SAD in liquid technique is extremely simple since it does not require neither a laser nor vacuum and it works at room temperature. To now, a great number of organic solvents has been employed (e.g., acetonitrile, decalin, toluene, n-hexane, n-dodecane, mixture of acetonitrile and water) showing production yield depending on the specific solvent. The use of water is appealing for low cost and environment-friendly applications. Indeed, an early attempt has been reported in literature concerning water as a solvent to show that H termination was provided by the solvent itself [23]. In that work the maximum detected polyyne length in the mixture was up to 10 carbon atoms, no detailed analysis on the process parameters has been provided nor the collection of size-selected samples was attempted.

The investigation of polyynes can be performed by UV-Vis and Raman spectroscopy. In fact, the size-dependent electronic structure leads to a sequence of vibronic bands in UV-Vis spectra which shifts to higher wavelength by increasing the number of carbon atoms [24, 25]. Vibrational properties can be investigated by Raman spectroscopy, showing peaks located in the 1800-2300 $cm^{-1}$ spectral range as a fingerprint of sp-carbon [26]. The red-shift of the Raman peaks with increasing the wire length can be used to characterise the size distribution [27, 28]. In case of weak Raman signal of isolated sp-carbon chains, or when the polyynes concentration in solution is too low, surface-enhanced Raman scattering (SERS) has proven to enhance the sensitivity up to 6 order of magnitude by exploiting silver in form of nanoislands supported on surfaces or colloidal nanoparticles [29, 30].

In this paper we present the submerged arc discharge (SAD) in water as a simple, cheap, and environmental-friendly technique for the synthesis of hydrogen terminated polyynes H-$C_n$-H (just $C_n$ for simplicity in the following). The optimisation of the process parameters allowed us to produce hydrogen-terminated polyynes up to 16 atoms of carbon, comparable to the ones produced with organic solvents. The produced polyynes were structurally characterised as a mixture and as size-selected samples after separation by high-performance liquid chromatography (HPLC). Size-selected polyynes showed longer stability in liquid with respect to the polyynes mixture under the same experimental conditions, while stability after solvent evaporation can be achieved by using Ag colloids used for SERS investigations.

The use of water in place of flammable organic solvents open to the facile synthesis of polyynes which is a key step towards the possible employment of these systems in applications, as recently shown in the case of biosensors [31].

## 2. Experimental details

Polyynes were synthesised by means of submerged arc discharge in liquid using graphite electrodes with a purity of 99.999%, 5 mm in diameter and 200 mm in length (Testbourne Ltd.) The submerged arc discharge takes place in a water-cooled jacketed reactor filled with 60-100 ml of deionised water Milli-Q (0.055 µS) at room temperature. The two graphite electrodes are connected to a DC power supply (EA Elektro-Automatik model EA-PSI 9080-60 T (0–80 V and 0–60A)) and arranged in "V" geometry.

The concentration of polyynes in the mixture was estimated by measuring UV-Vis absorbance with a Shimadzu UV-1800 UV/Visible scanning spectrophotometer (190-1100 nm). High-performance liquid chromatography HPLC was performed by a Shimadzu Prominence UFLC equipped with a photodiode array UV-Vis spectrometer, i.e. diode array detector (DAD) and a Shimadzu Shim-pack HPLC packed column GIST C8 (250 mm x 4.6 mm, 5 µm particle size). All the solvents were purchased by Sigma-Aldrich with HPLC grade.

The solution containing the mixture of polyynes was filtered through polytetrafluoroethylene (PTFE) syringe filters (25 mm diameter) having pore size of 0.45 µm in order to avoid contamination for HPLC. For the preparative procedure we used the so-called batch-loading on column method where crude polyynes mixture in water was accumulated on HPLC column to favour the attachment onto functionalised silica particles.

HPLC was operated both in isocratic and gradient mode with the mobile phase pumped at a flow rate of 1 ml/min into the HPLC column. In isocratic mode the mobile phase consists in acetonitrile/water 80/20 v/v, whereas in gradient mode the mobile phase starts with acetonitrile/water 5/95 v/v and reaches acetonitrile/water 100/0 v/v after 35 min of operation.

Structural characterisation was carried out by UV-Vis absorbance spectroscopy and by Surface-enhanced Raman Spectroscopy (SERS) both in solution and in solid-state on SERS active substrates.

Silver colloids ($10^{-3}$ M) made by Lee-Meisel method [32], with a size of 60–80 nm of diameter (from SEM image) and showing a plasmonic peak at about 413 nm, were added to the polyynes solution in water 50/50 v/v in order to perform SERS analysis in solution. SERS active substrates made of Ag nanoparticles on silicon were evaporated by Edwards E306A evaporation system, following the procedure reported elsewhere [33]. Solid-state SERS is performed by drop casting the solution of polyynes on the SERS active substrate and SERS spectra were acquired using a Renishaw InVia micro Raman spectrometer (spectral resolution of about 3 cm$^{-1}$) with a laser power of 1 mW at the sample from an argon ion laser ($\lambda$=514.5 nm).

First-principles calculations of the vibronic spectra of $C_n$ were carried out by employing time-dependent DFT (TDDFT) calculations, adopting the Coulomb-attenuated range separated exchange-correlation functional CAM-B3LYP, and the Dunning's correlation consistent cc-pVTZ basis set. All calculations were performed by using the package Gaussian09 [34, 35]. For each $C_n$, after the ground state geometry optimisation, vertical excited state energies were computed at TDDFT level. For the most relevant dipole-active state (mainly described by single-particle HOMO-LUMO transition) a further TDDFT geometry optimisation and force field calculation have been performed. Based on ground and excited state equilibrium geometries and vibrational force fields, the Huang-Rhys factors and the vibronic spectra were evaluated, as reported in Ref. [35].

### 3. Results and Discussion

*3.1 Optimisation of the process parameters*

Polyynes were produced in water by arcing graphite electrodes at room temperature. The solution of water/polyyne after the arc discharge resulted rich in carbon particulate (inset in ***Figure 1)***, which is the reason why we filtered it before all the characterisation measurements. Several experiments were performed to optimise process parameters (i.e. discharge current, arcing time and depth of electrodes). UV-Vis spectra of aqueous solution right after the discharge and filtration were carried out to easily detect the presence of polyynes. H-terminated polyynes are more likely to be soluble in organic solvents than in water, even though no detailed analysis is present in the literature about their solubility. We observed no precipitation in the solution after filtration, thus indicating possible solvation of polyynes in water maybe due to their low concentration, as noticed also by Compagnini and co-workers [36, 37].

All the spectra reported in *Figure 1* present the same peaks at 199, 206, 215, 225 nm, associated to polyynes of different length, as discussed below. The role of discharge current, arcing time was investigated by comparing the intensity of UV-Vis peaks, estimated by the peak-valley difference in absorbance. In particular, we considered the peak located at 199 nm, which is attributed to $C_6$ and the peaks at 206, 215 and 225 nm attributed to $C_8$ (see **Figure 1**).

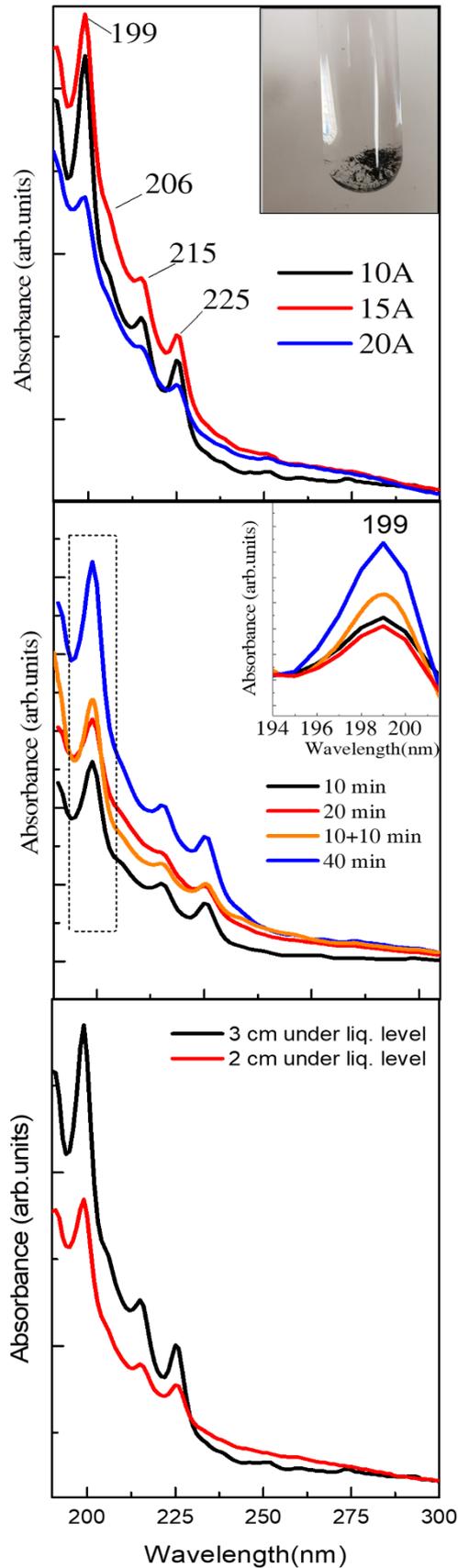

***Figure 1.*** *Comparison between UV-Vis spectra at different working parameters. **a**) Current. **Inset:** solution of polyynes in water after the discharge before filtering. **b**) Time of arcing. **Inset:** Detail on the peak-valley at 199 nm. **c**) Depth under the surface liquid level at which the electrodes form the arc discharge.*

To obtain a good discharge with current less than 10 A, a voltage of at least 25 V was required. Three different experiments were performed changing the current in the range 10-20 A at fixed voltage of 25 V and with arcing time of 10 minutes. The peak-valley difference at 199 nm is larger at 10 A (24% more than the one at 15 A and 87% more than the one at 20 A) and the same trend was found at 225 nm with values of 32% and 78% larger, respectively. This shows a larger production of polyynes for smaller current values.

In literature, a few works showed an increase of the concentration of polyynes with increasing duration of the discharge up to 150 s, maintaining fixed current, voltage and water volume [21]. Here, by extending the time of the process to 10, 20 and 40 minutes we observed a more complicated behaviour. In fact, the intensity of UV-Vis peaks decreases after 20 minutes of continuous discharge compared to 10 minutes, and it increases again after 40 minutes. The latter observation is clearly visible from the inset of *Figure 1b*, after matching all the valleys related to the peaks at 199 nm. A possible explanation for the decrease in the absorbance after 20 minutes is that the arc discharge during the synthesis of polyynes is detrimental for already formed species, due to UV radiations and heat generated in the discharge which can affect polyynes stability. Moreover, it is reasonable that, by increasing the arcing time, polyynes concentration reaches a ~~maximum~~ value at which degradation mechanisms (mainly cross-linking reactions and photodegradation) become relevant, especially for the longest species.

To overcome this limitation, we included breaks of 5 minutes every 10 minutes of discharge. This allowed to cool down the solution from 41 °C to the initial temperature of 23°C and allowed polyynes to spread out from the plasma region. In this way, exploiting breaks, at comparable arcing time, we obtained higher polyynes concentration. The drawback of this strategy is the longer duration of the synthesis which increases possible degradation process due to exposure to air. Therefore, we observed two coexisting effects during the synthesis of polyynes, the detrimental action of the discharge for already formed polyynes in the plasma region and the diffusion of polyynes in the solvent. 40 minutes of arcing is possibly enough to allow diffusion of already formed polyynes in all the solvent volume, justifying the higher concentration.

We observed that for arcing time in the ten minutes range the concentration does not increase linearly with time, hence in the following investigation we fixed the arcing time at 10 minutes.

We also observed that the depth of the electrodes under the solvent surface plays a role in the formation process. In fact, when the electrodes are less immersed in the liquid (2 cm below the liquid level with respect to 3 cm), the plasma formation is closer to the liquid surface level and we observed an increase in the liquid temperature from 40°C up to 80°C, as measured at the external edge of the reactor. Moreover, close to the liquid surface, the plasma is less confined and becomes more unstable as indicated by the larger quantity of produced bubbles. All these factors may lead to a lower concentration, also considering that polyynes are constrained to stay closer to the plasma zone and more subjected to UV radiation and heat. Based on these results we chose 10 A, 25 V in 100 ml of water for 10 minutes as optimal parameters for the detailed characterisation of polyynes produced by SAD in water, as discussed in the following.

*3.2 Characterisation of size-selected hydrogen-capped polyynes*

After UV-Vis characterisations of water/polyynes solutions for the selection of good process parameters, HPLC allowed us to concentrate and to separate polyynes in size-selected samples.
To enable HPLC separation, we need a mobile phase composed of a mixture of organic solvent (i.e. acetonitrile) and water, as detailed in the experimental section. Polyynes are more soluble in acetonitrile than in water and we noticed the increase of their solubility with the decrease of the length of the chains. In this way, the shorter chains are less retained on the stationary phase and will exit from the chromatographic system first.
*Figure 2* shows the UV-Vis absorption spectra obtained by the DAD coupled with HPLC of size-selected hydrogen-capped polyynes $C_n$ (n=6-16) synthesised in water. Spectra have been normalised for better comparison even though the intensity decreases up to three order of magnitude for longer polyynes. The observed absorption bands are related to the vibronic progression mainly governed by the $C \equiv C$ stretching mode. The absorption energy decreases to longer wavelengths by increasing chain length, in agreement with the spectra of hydrogen-capped polyynes as reported in literature [29][38].

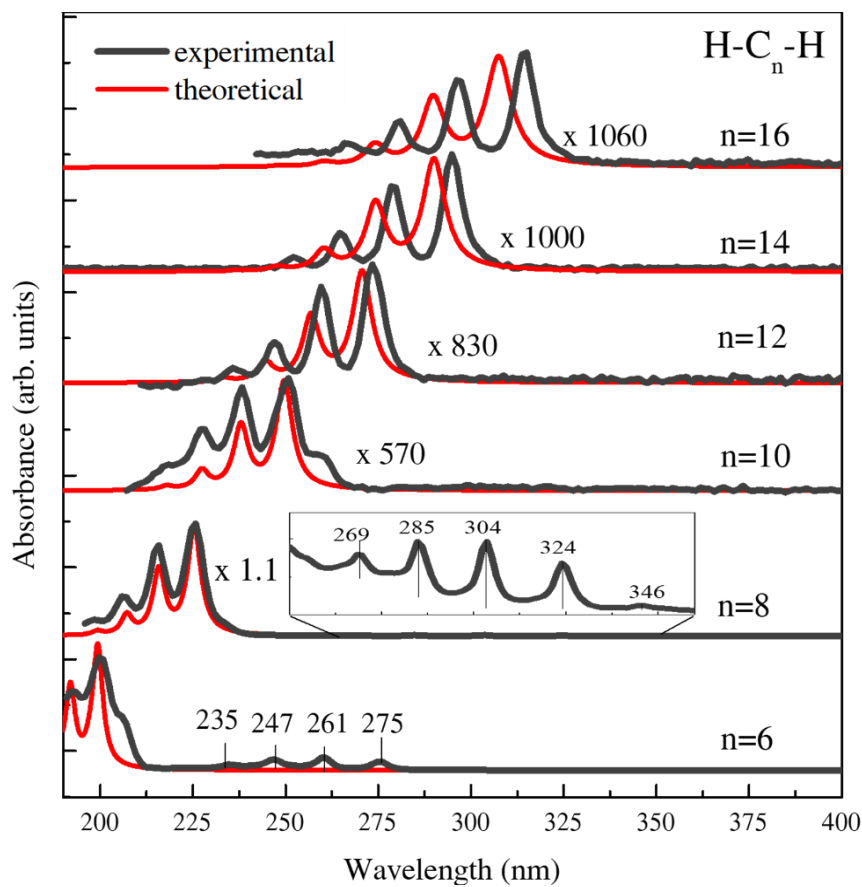

*Figure 2. In black, normalised absorption spectra of hydrogen-capped polyynes $C_n$ (n=6-16) in water separated and collected by HPLC. In red, DFT calculations of UV-Vis spectra compared with experimental data. The inset shows the low intensity peaks of $C_8$ magnified.*

We identify H-capped polyynes up to n=16 carbon atoms (i.e. $C_{16}$) synthesised in water. In this way, after optimisation of the process parameters and ad hoc concentration procedure on HPLC for aqueous solution, we were able to analyse longer polyynes with respect to a previous report using water as a solvent in the arcing process (see e.g. Ref. [23]).

In case of $C_6$, due to the cut-off of the UV-Vis spectrometer at 190 nm, we can observe only one intense vibronic peak at 199 nm together with low intensity peaks at 235, 247, 261, 275 nm, related to forbidden transitions [21, 39, 40]. Also for $C_8$ we notice forbidden transition as low intensity vibronic peaks at 269, 285, 304, 324, 346 nm, with small differences compared to Ref. [24]. Such low intensity peaks are not detectable for longer polyynes, probably for the weak intensity of the signal.

To support the interpretation of our UV-Vis results we performed TDDFT to compute vibronic spectra (see *Figure 2*), which are in good agreement with experimental data. All the predicted spectra have been rigidly shifted by 19.5 nm to match the experimental data of $C_8$. Indeed, $C_8$ has been widely studied in previous works [18, 30, 41, 42] and many spectra are available for a comparison, showing a complete agreement with our spectrum. For this reason, we considered it as a reference system for a reliable determination of the shifting factor applied to the computed spectra. Moreover, contrary to $C_6$, the UV/Vis spectrum of $C_8$ lies completely in the available experimental range.

Transition energies are well described at the TDDFT level (after the scaling procedure), nevertheless a little discrepancy with respect to the experimental values still remains. The energy overestimation can be inferred to a partial multireference character of the polyyne excited states, not considered by TDDFT methods, which would affect the computed excitation energy.

The vibronic progression of each $C_n$ is mainly dominated by one vibrational normal mode, attributed to the in-phase C=C/C-C stretching/shrinking of the carbon chain and which corresponds also to the most intense Raman active vibration. These results show that TDDFT calculations allow to predict vibronic spectra of polyynes as they can be experimentally detected by HPLC thus opening to the investigation of novel polyyne systems.

From the absorbance obtained by UV/Visible scanning spectrophotometer measurements and using the Lambert-Beer law with molar extinction coefficients taken from literature [24, 25], we estimated the relative fractions of polyynes present in the solution of water/polyynes obtained just after the synthesis and filtration. In case of $C_6$, we considered the molar extinction coefficients related to low intensity peaks ($\varepsilon$ = 100-350 L mol$^{-1}$cm$^{-1}$) and we found a value of about $10^{-6}$ M (93,2% as relative molar ratio) while for $C_8$ we obtained a concentration in the order of $10^{-7}$ M (6,4% as relative molar ratio) and for $C_{10}$ in the order of $10^{-8}$ M (0.4% as relative molar ratio ratio). The fraction of longer polyynes was difficult to estimate from UV-Vis spectra due to the decreasing concentration with increasing the polyynes length, anyway, based on DAD spectra, we expect a concentration less than $10^{-9}$ M.

After HPLC separation of H-polyynes, we were able to collect and to characterise only the most concentrated ones as size selected polyynes (i.e. $C_6$, $C_8$, $C_{10}$). The concentrations of size selected polyynes were so small to hinder the Raman signal of polyynes, hence we performed SERS measurements.

SERS spectra in liquid were acquired at 514.5 nm excitation wavelength by adding Ag colloids to the size-selected polyyne solutions. Ag colloids show a plasmonic peak at about 410 nm, however a broadening of the absorption spectrum due to aggregation of polyynes with Ag colloids allows absorption even at the laser excitation wavelength. In fact, after mixing Ag colloids with the filtered aqueous solution of polyynes after the synthesis, a shoulder at high wavelengths of the plasmonic band appears, as depicted in *Figure 3*. This allows to have SERS enhancement at 514.5 nm, in agreement with previous results [29, 33, 41, 43].

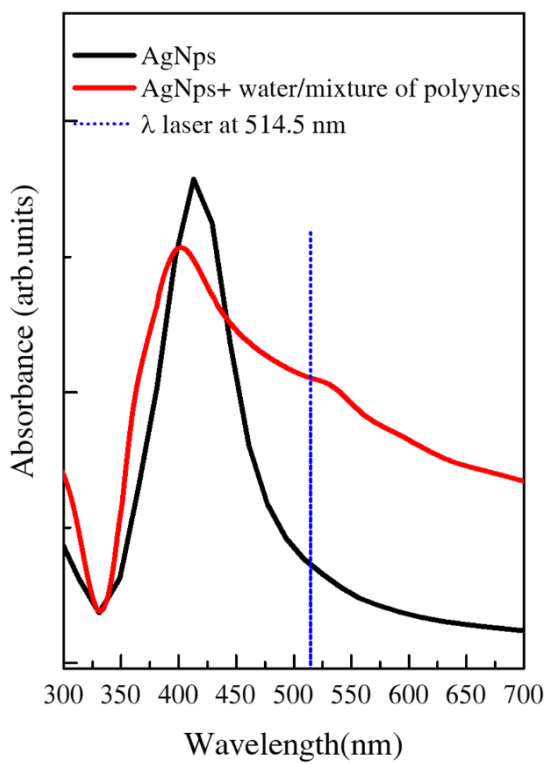

*Figure 3. Absorption spectra of as-prepared Ag colloid (in black) and Ag colloid with filtered mixture of polyynes in water (in red). Dotted line indicates the λ of the Raman laser at 514.5 nm (in blue).*

The SERS spectra of size-selected samples, i.e. $C_6$, $C_8$ and $C_{10}$, were carried out in order to assign the origin of the bands in SERS spectrum of polyynes mixture. Acetonitrile is present in the collected fractions of $C_6$, $C_8$ and $C_{10}$ because they were obtained after separation on HPLC, in which the mobile phase (acetonitrile/water) flows together with the liquid sample. For this reason, we added the same quantity of acetonitrile also to the water solution of polyynes mixture in order to analyse the samples at the same experimental conditions. The $sp^2$ signal was negligible and only the sp region is reported in *Figure 4*.

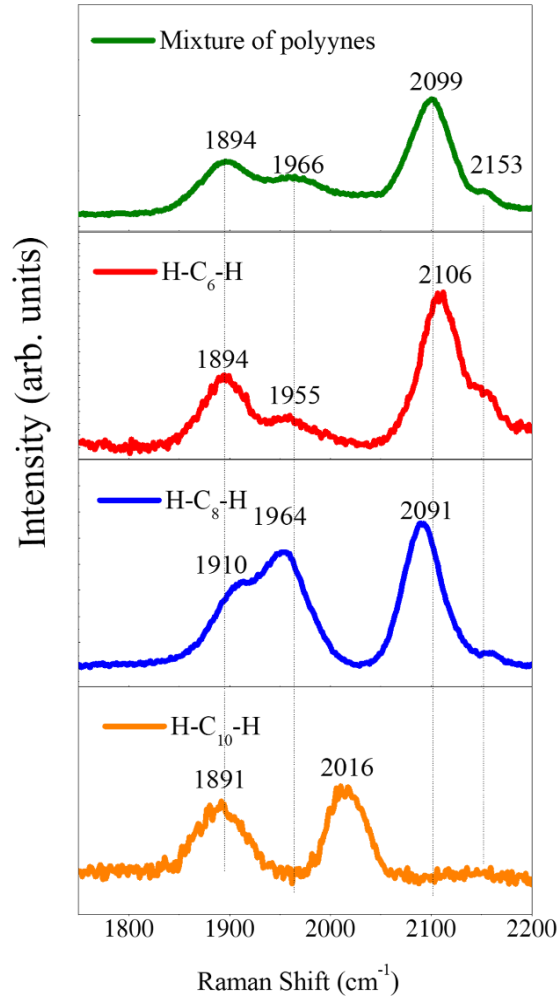

*Figure 4. SERS spectra of **a**) the polyynes mixture, **b**) size-selected polyynes with n=6, **c**) n=8 and **d**) n=10.*

All the spectra are composed of two bands: one in the 2000-2100 cm$^{-1}$ range (also called α' band in Ref. [30]), typical of polyyne-like chains and one in the range 1800-1900 cm$^{-1}$ (also called β' band in Ref. [30]), which usually appears in SERS of polyynes, in agreement with previous works [29, 44]. Here, the SERS spectrum of polyynes mixture appears as the superposition of the spectra of size selected samples, considering the different concentrations and the Raman intensity of the different polyynes lengths. In fact, the Raman effect depends not only on the concentration of the species but increases in intensity with the length of the chain [29].

Our results are in good agreement with SERS spectra of H-capped polyynes reported in literature [29, 30] and seem to follow the work by Hanamura et al. regarding an observed red shift with the decrease of Ag nanoparticles size [41]. By changing the nanoparticles size from several hundreds of nm (i.e. Ag islands) as in Ref. [30, 41] to less than 20 nm the peak frequency moves toward higher wavenumber [41].

By employing Ag nanoparticles of 60–80 nm we obtain peak frequency between the two conditions mentioned above. Further studies need to be done in order to better understand possible mechanisms responsible for shifts in the vibrational frequencies due to the size and shape of Ag nanoparticles.

*3.3 Stability investigation*

We studied the stability in time of hydrogen-terminated polyynes at different experimental conditions in order to find strategies for their long-term stabilisation. UV-Vis spectra of polyynes mixture at different ageing times are reported in *Figure 5*.

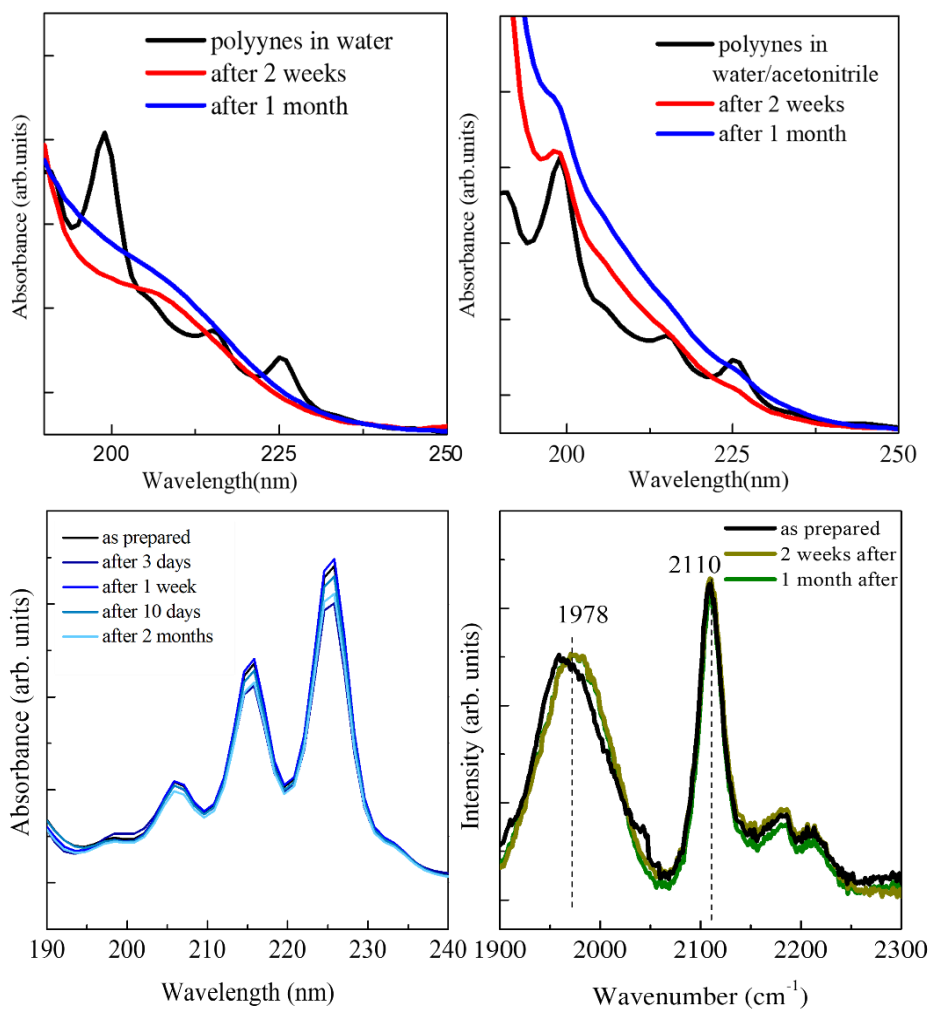

*Figure 5. Stability investigation: a) UV-Vis spectra of polyynes mixture in time after dilution with water, b) UV-Vis spectra of polyynes mixture in time after dilution with acetonitrile, c) UV-Vis spectra of $C_8$ polyynes at different ageing time and d) SERS spectra of dried polyynes mixture with Ag colloid.*

The filtered solution of polyynes mixture prepared by SAD in water was divided in two samples and both diluted with the same amount of solvent 1/5 v/v%, one with water and the other one with acetonitrile. The two samples were studied in time thanks to UV-Vis spectroscopy. After 2 weeks and even more after 1 month, we noticed a degradation in both cases but with acetonitrile we can still identify some features of the peaks of polyynes, see *Figure 5a e b*. The beneficial effect of acetonitrile was also observed by Cataldo [45]. Explanation of this phenomenon can be associated to the solubility of polyynes, in fact polyynes are more soluble in acetonitrile than water, so this can help in avoiding aggregation and hence cross-linking reactions. Moreover, the higher quantity of oxygen dissolved in water in respect of the acetonitrile can boost the oxidation process of polyynes.

Another experiment which confirmed that acetonitrile can help in stabilizing H-polyynes was related to the study of ageing effects of a collected fraction of size-selected polyynes, i.e. $C_8$, after HPLC separation and recording UV-Vis spectra as a function of time from DAD (Figure 5c). After 2 months the signal is almost unchanged. This result can be also associated to the process of separation itself, the low quantity of air inside the closed vial and the lower concentration of polyynes present in the solution. In fact, in a liquid sample, cross-linking reactions are responsible for the polyynes degradation and are expected to depend on the concentration [12].

The collection of fractions of size-selected samples after HPLC separation appears as a way to improve stability, even though the precise mechanisms need to be investigated with further experiments.

Furthermore, prolonged stability was observed even in SERS spectra of dried polyynes mixture with silver colloid on a silicon substrate over time (Figure 5d). The shape of the sp-carbon signal is different from the SERS spectra of the solutions but reveals the presence of the main band at 2110 cm$^{-1}$. The bands at 1978, 2184 and 2215 cm$^{-1}$ do not overlap with any bands of in-solution SERS spectra previously reported, showing that SERS effect strongly depends on the interaction between the analyte and the colloid. We observe that the signal of polyynes does not change after 1 month, in accordance with previous works [46], confirming that this method allows to stabilise the polyynes mixture independently from the solvent where they are produced. In fact, polyynes bonded to silver are probably constrained in a certain position avoiding interactions and cross-linking.

## 4. Conclusions

We have shown the use of pure water for the synthesis of H-capped polyynes by means of the submerged arc discharge in liquid. After optimisation of the process parameters and ad hoc concentration procedure, polyynes ($C_nH_2$) from 6 to 16 atoms of carbon were identified and separated by HPLC. The recognition of all the polyynes through UV-Vis spectroscopy from DAD was confirmed by TDDFT calculations, showing the capability of producing long polyynes in water as in case of organic solvents. Stability of hydrogen-terminated polyynes was also studied comparing spectra of different samples: water/polyynes, acetonitrile/water/polyynes acetonitrile/water/size-separated polyynes ($C_8$) and dried Ag colloid/polyynes on a substrate. The last two mentioned samples showed a stability exceeding 1 month. Our investigation has been aimed at optimizing simple, low-cost, and easy to scale-up techniques to produce sp-carbon chains, as SADL. In this perspective, the use of water as a solvent allows to form hydrogen-capped polyynes similar, in terms of length, to the case of organic liquids, for which a wide amount of data is available. This enables us to consider these polyynes as a good benchmark for improving the knowledge of sp-carbon chains, even if their limited stability does not make them the best candidates for future applications in absence of further strategies (i.e. embedding them in a protective matrix or in nanocomposites). In any case, our results suggest the promising applicability of SAD in liquid in view of possible mass production, paving the way to future investigations on the optimisation of this technique to produce more stable sp-carbon systems.

## 5. Acknowledgements

Authors acknowledge funding from the European Research Council (ERC) under the European Union's Horizon 2020 research and innovation program ERC-Consolidator Grant (ERC CoG 2016 EspLORE grant agreement No. 724610, website: www.esplore.polimi.it).
Authors also acknowledge Andrea Dalla Via for the preliminary work done in his Master thesis.